# Coddlers, Scientists, Adventurers, and Opportunists: Personas to Inform Online Health Community Development

Jina Huh[1], Bum Chul Kwon[2], Jaegul Choo[3], Sung-Hee Kim[4], Ji Soo Yi[5]
Michigan State University[1]; University of Konstanz[2]; Korea University[3]; University of British Columbia[4]; Purdue University[5]

**Abstract**

*As online health communities (OHCs) grow, users find it challenging to properly search, read, and contribute to the community because of its overwhelming content. Our goal is to understand OHC users' needs and requirements for better delivering large-scale OHC content. We interviewed 14 OHC users with interests in diabetes to investigate their attitudes and needs towards using OHCs and 2 OHC administrators to assess our findings. Four personas— Coddlers, Scientists, Adventurers, and Opportunists—emerged, which inform users' interaction behavior and attitudes with OHCs. An individual can possess the characteristics of multiple personas, which can also change over time. Our personas uniquely describe users' OHC participation intertwined with illness contexts compared to existing social types in general online communities. We discuss broader implications back to the literature and how our findings apply to other illness contexts in OHCs. We end with requirements for personalized delivery of large-scale OHC content.*

**Introduction**

Studies showed OHCs help patients improve both psychosocial and behavioral aspects in managing health[1,2]. However, the abundance of information provided by OHCs can be challenging for users to properly digest and interact with the communities. Studies of online communities have used discussion interfaces[3], text analytics[4], and visualizations[5] to help users digest large-scale online conversations. Many of these techniques could be used for OHCs because of their similar interaction patterns. Online community participation patterns, such as lurking, moderating, and experiencing legitimate peripheral participation as newcomers[6], are also observed in OHCs. Gaining others' experiences on medical treatments in OHCs is analogous to getting product reviews from online communities on electronic gadgets, for instance[7].

The unique characteristic of OHC users, however, is that each individual has complex, serious—sometimes life threatening, health-related needs coming from various angles of their lives. Examples include personal life, stages of illness (e.g., newly diagnosed vs. settled in managing illness), complications, and preferences and philosophies toward approaching the illness. These personal contexts produce unique and complex user needs as such needs mesh with approaches towards interacting with OHCs. Thus, we do not know how much of the existing ways of efficiently delivering large-scale online discussions can be perceived as helpful for OHC users.

In this paper, we present interview results with 14 OHC users at various stages of illness and community use. We also discussed our findings with one administrator of a large OHC to understand how much the findings concur with his own experience overseeing his OHCs. Our findings helped to identify: (1) personas representing varying user needs in their interactions with OHCs and (2) requirements for dealing with large-scale OHC content delivery. We will discuss broader implications back to online communities, information behavior literature, and the implications of our results for OHCs in other illness contexts.

**Background**

*Information overload in generic online communities vs. OHCs*

Information overload in online communities is an ongoing problem researchers have long investigated. Researchers developed discussion interfaces, such as Conversation Thumbnail[8], which provided the context of overall discussion threads. The Looms project helped to identify thread structure and emotional content[9]. Text analytics projects examined ways to generate threads' summaries[10] and cluster related posts[11]. Researchers of online communities continued to work on personalization in digesting online conversations. Examples include: individualizing filtering preferences for reading comments on Slashdot[12] and recommending conversations from Twitter and Facebook[13].

Users of OHCs also deal with information overload. However, patients come to OHCs with somewhat distinctive motivations from general online community users. Users' complex, serious—sometimes life threatening—situations around managing illness are the central distinctions between general online community and OHC. Aforementioned studies regarding delivering online conversations have great potentials to be applied to OHCs. However, we still do not have answers to the question of how much techniques tested under the context of general online communities can be helpful in the health context. Questions arise as to whether such techniques for solving information overload will even be perceived as helpful for OHC users. OHC users might prefer to read all threads despite the inefficiency. OHC users might have other needs not addressed from generic online community interfaces.

This "other need" can be psychosocial support. The literature on OHCs emphasized the crucial role of social support exchanged among the community users[14]. The self-help groups, such as Alcoholics Anonymous®, are established based on the fact that patients want to help others, which helps themselves as well[15]. Numerous studies showed how psychosocial help in OHCs led to successful illness management[2,16,17]. Interacting with peers as mentors significantly improved patients' diabetes conditions compared to receiving help from the nurse navigators[18].

Still, not only psychosocial support, but also informational support plays a critical role in OHCs. The social support theory categorized both emotional and informational support as core constructs under social support[19]. Studies have shown that patients visit OHCs to ask questions on the illness, gain peers' expertise, and educate themselves about daily management strategies[20].

As such, OHC users can develop multifaceted needs, wanting both psychosocial and informational support. Furthermore, each individual with diversified needs due to varying illnesses, complications, personal contexts, and learning styles will all have to be taken into account. Thus, the advanced interfaces developed towards large-scale online conversations might only partially address the OHC users' needs because these interfaces were not intended to support such complex illness contexts. Whether users will perceive current OHCs as having information overload problems is also questionable. Identifying what user types and needs there are would be an important next step.

*Social types in online communities vs. OHCs*

Researchers developed a number of social types and roles online. Fisher et al.[21] divided online community users into two categories—information providers and information users. Golder[22] described newbie, celebrity, elder, lurker, flamer, troll, and ranter. Kim[23] described how social types change over time as their participation with the communities increase: from visitors to novices, regulars, leaders, and elders. Turner et al.[24] described: answer person, questioner, troll, spammer, binary poster, flame warrior and conversationalist. All of these social types were defined by users' posting and participation behavior.

The distinctive nature of OHC users—psychosocial issues involved in illness management—adds complexity to how social types can be categorized. In addition to the personal context around having the illness, privacy issues, information quality, and legality all emerge as new issues in understanding OHC users. Thus, posting and participation behavior should not be the only measure in which social typing occurs. Such existing models of social types and roles in online communities fall short in addressing psychosocial and informational aspects strongly tied to illness management. We should first understand what personas there might be, incorporating all aspects of information behavior—emotional, interpersonal, and psychosocial characteristics in users' interactions with OHCs. Then we can think about re-appropriating existing or new technologies to present large-scale online conversations.

**Methods**

Our interviews were in two parts. First, we conducted a total of 14 interviews with OHC users who are either diagnosed with diabetes (both type I and II) or have concerns with diabetes as their health interests. We then interviewed 2 OHC administrators to confirm our interview findings with the 14 OHC users, given his long experience of overseeing multiple online health communities. For the first interviews, we chose diabetes because it is a chronic illness where they can benefit from both emotional and informational support[25]. We wanted to recruit participants at all stages of managing diabetes and OHC use. Accordingly, we included those concerned with diabetes but did not have the diagnosis in the participation criteria and those with varying levels of experience using OHCs. The recruiting sites were diabetes-related online communities and subgroups (3 online diabetes communities, reddit.com, and a diabetes Facebook group) and co-authors' social network (e.g., friend's friend). The age of the participants varied from 19 to late 60s. OHC user participants included 8 females and 6 males, 1 type I diabetes, 8 type II diabetes, and 5 undiagnosed but have concerns with diabetes. The OHC administrator was male and a caregiver of a diabetes patient. We recruited 6 from online diabetes communities, 5 from reddit.com, 1 from the Facebook group, and 3, including the one OHC administrator, from our social network. We emailed a number of OHC administrators found online and one administrator responded to agree participating in an interview, totaling our interview pool to 2 OHC administrators in addition to 14 OHC users. Our OHC user participants' number of years since diagnosis varied from undiagnosed to 2 weeks to 20 years. Also, their experience with OHCs varied from one-time users, lurkers, to regular users. Our study has been approved by the Institutional Review Board.

**Table 1. Characteristics of the personas in online health communities**

|  | Coddlers | Scientists | Adventurers | Opportunists |
|---|---|---|---|---|
| **Illness characteristic** | Experienced patient | Transitioning patient | Exploring patient | Newly diagnosed |
| **Participation characteristic** | Posts emotional support messages | Moderates or Researches while lurking | Shares useful information or observes over time | Lands on OHC through search results |
| **Consumption characteristic** | Reads all threads; relies on others for quality assessment | Needs scientific evidence | Looks for unusual, unordinary information | Looks for straight up answers |
| **Values sought from OHC** | Sense of community | Relevant, validated experiences | Avoiding mainstream | Reviews and experiences |

The semi-structured interviews took 1 to 1.5 hours for each participant. We asked OHC user participants to share their experiences and concerns around diabetes and using OHCs. We then gave them a webpage we created, which contained 206 diabetes community threads on Atkins diet downloaded from a publicly accessible online diabetes community. We did so to remove distracting links and advertisements. We chose the Atkins diet as the topic because it helps us to probe on factual as well as experiential knowledge (e.g., What were the experiences of Atkins diet users?), which is a representative online community content. We asked participants to find what they want to know further about Atkins diet as if they are reading through community threads. During this process, we asked probing questions to further understand their choices of what to read and why. For the interview with the OHC administrators, we asked them what types of patients they like to support in their communities. We then probed what different patients there might be and how they attempt to support them differently as administrators. We then shared our results to check whether the findings concur with their understanding of the OHC user types in their OHCs.

All interviews were transcribed. Using open coding analysis[26], all authors first analyzed one participant's transcript together. We then shared our thoughts and initial codes. We negotiated and merged our codes, with which we further coded the rest of the interviews while allowing for new codes to be developed. For any one participant's transcript, at least two coders were present. Through the affinity diagramming exercise[27], we identified emerging themes, particularly around characterizing OHC use and how these usage behaviors further group into personas.

**Personas: participation, consumption, and values sought in online health communities**

From the affinity diagramming exercise, four distinctive user patterns of participating, consuming, and value seeking behaviors toward OHC emerged. We labeled these lumps of patterns as personas—Coddlers, Scientists, Adventurers, and Opportunists, as summarized in Table 1. At any given time, one individual can have characteristics of more than one personas, although there can be one or two dominant persona(s). This individual can also move from having one persona to another over time as their personal contexts and needs change.

*Coddlers: Experienced patients--OHCs as a social space*

For Coddlers, OHCs are a social space where they altruistically engage by helping others and maintaining emotional bond with peer users. Accordingly, visiting OHC is part of their daily routine. They have been participating in the community for many years, and they gain value from interacting with others and exchanging encouragements and stories as an experienced diabetes patient.

Participation: Community building through regular, loyal, and altruistic participation. Coddlers are regular visitors of OHCs, mostly loyal to a single OHC. They tend to be old-timers. Accordingly, they tend to be more experienced patients with diabetes than other members, where their diabetes regimen are settled. Thus, they now have the ability to help others, given what they have learned from their own years-long experiences. These participants developed a sense of ownership with the communities, referring to the communities as "we" (P3). Accordingly, when Coddlers read conversation threads, they find places to help whenever they can and attempt to help with newcomers and others who tend to be new to the community as well as in managing diabetes: *Yeah, I usually read all the responses, you know, especially in the new questions. And then, [see] if I feel like I can contribute* (P3); *[I would say] 'Yes, it happens, pick yourself up, move on.' And hopefully, I help other people.* (P13). Furthermore, P13 stated how helping others "also helps himself," emphasizing the benefit of mutual help exchange in OHCs.

Coddlers' goal in participating in OHCs is less about practical information seeking, but more about community building. They consider posting as a way to keep up with the community and help others. For instance, P1 said that posting not just about diabetes but also about "the personal side of things" can help build a community together. Because Coddlers' primary goal in responding to threads is interacting with other community members, when choosing which threads to read and respond to, how recent a post was posted is important. In this manner, Coddlers can have a better chance of getting a response back from the posters of the thread (P3). Not only when posting in the communities, but also when consuming information, Coddlers engage relating with other community members.

Consumption: Getting into the weeds while allowing accidental information discovery. As regulars of the community, Coddlers rarely perform focused search for a particular subject. Coddlers' visiting routines mostly consist of checking for new threads or new comments since their last visit. They like to discover new information as they browse through the threads. They rely on other community members to assess the quality of the information posted, rather than looking for scientific evidence.

Accordingly, they read or at least skim all unread threads. During this process, they would often "get into the weeds," (P2) while attempting to fully understand the context of the conversation (P10). The participants who showed this particular characteristic of a Coddler—reading all threads instead of selectively choosing what to read—found original threads with all replies to be more useful than any summarized version of the threads (P10). The nuance would be lost in how posters portray their personal situations and opinions if the threads were somehow summarized or edited with an automated process (e.g., showing extracted quotes or aggregated responses).

Coddlers consider visiting OHCs as part of a daily routine, socializing with other community members and gaining additional information as a bonus. Thus, when reading the threads, Coddlers go with the flow, consuming information as they arise. P3 expressed how her needs at the moment influence her to choose specific posts in further detail. For P10, he would find useful information unexpectedly as he reads all comments in detail, such as finding out about stressing muscles during exercise: *I had started exercises using a stationary bike and just wasn't paying attention that I was stressing my muscles out that hadn't been working for years. [chuckle] It was just good information for me and I was happy to utilize it and use it* (P10).

When Coddlers want to evaluate the information quality of a post, they would rely on other community members' confirmations or opinions about the validity of the post: *"Oh, it happens to me, the same thing." If I see more people confirming the earlier posts, then I start to have some kind of comfortable feeling* (P10). Similarly, P8 would check on people's comments before visiting the posted link to evaluate the content.

Coddlers trust their community members' opinions and consider a forum as more reliable than other information resources because of the interactions with others (P9). Coddlers consider OHCs sufficient as an information source. OHCs provide "everything under the sun": *There's recipes, discussions, just about everything under the sun, that I probably had wondered about when I was early diagnosed, that I wished I had known then* (P10).

Values sought: The sense of community. The sense of community and emotional support is the crucial value that Coddlers believe to be the key in OHCs. Coddlers prefer that the members of the community do not have to face the reality in a harsh way, especially for newcomers, who tend to already have hard time dealing with a new diagnosis. Coddlers consider a post "useful" and "complete" if the poster adds encouragement to information: *Very useful post. First, it praises the person and encourages them about what they are actually doing. So it's giving them some motivation. [...] some posts would say, "Don't do that." But this is a complete post. They actually praised them, encouraged them about what they actually are doing, so they don't just make them feel down* (P9).

Coddlers bond with other members through real-time chatting or private messages (PM): *There is a new person that came on the forum not too long ago and it was obvious she was struggling [in getting diabetes under control]. [...] So, she was going by what the doctors have told her, what her family tells her, and what little bit she's gleaned from discussions. And I felt we've kind of bonded and we PM each other and we exchange a lot of information* (P1).

It is not just about exchanging information, but the chat becomes a personal conversation or emotional support (P3). P1 also agreed that members bond with each other further through other personal contexts outside the illness, exchanging irrelevant conversations to diabetes, such as, "Hey, you're from the west, so am I."

For Coddlers, even as an experienced diabetes patient, the emotional support and the sense of community are as important as helping members manage the disease: *I don't know. It feels good to just reach out to somebody who I have diabetes in common with. I like people so I don't want to be an island and isolate myself, because I feel like in the past, I've done that enough* (P1).

***Scientists: A transitioning patient--Skeptically, OHC as a place to gain information***
Scientists continue to test and look for strong evidence from the information shared around OHCs. Unlike Coddlers, whose main goal of participating in OHCs is to exchange emotional support, Scientists are looking specifically for information relevant to their personal health profile. Scientists tend to be at a transitioning stage of illness where they are looking for alternative regimen that can address a new challenge they faced or a new intriguing health management method they found. Thus, Scientists maintain a critical eye on choosing what information to consume.

Participation: Either an active moderator or a silent researcher. Depending on their level of participation, Scientists can be active moderators or silent researchers. One important characteristic that distinguishes Scientists from others is that they like to verify information shared in OHCs. Such a characteristic is helpful to have as a moderator, who monitor inappropriate information exchanges among members. Also, for those looking for a new regimen because they are having problems with an old one, gathering valid information on the new regimen would be critical. P2, with experiences in moderating general online forums, wanted to make sure that no misinformation is being shared, especially for information which does not pertain to every individual. P13, who also has been a moderator for multiple online forums, makes sure the community is free of any arguments or inappropriate posts: *Mainly I'm looking for any flagged posts, any arguments, people getting rude. Sometimes you get spammers* (P13).

Other participants with the characteristics of Scientists showed an opposite behavior in terms of community involvement. Since Scientists believe that they should have an actual experience or evidence of a success or failure in managing diabetes to post anything, they remained as silent researchers for many years as P8 did: *I don't really feel like I've made any weight loss progress, so I don't really feel like I have reason to make a post* (P8).

Consumption: Information scent for judging quality and relevance. Unlike Coddlers who trust community members' judgments on information quality, Scientists are skeptical. Scientists look for all possible indicators on how valid the shared information is. Scientists search for specific information. Thus, they do not embrace, like Coddlers do, irrelevant conversations, such as sidetracked conversations or personal chats. Some participants were skeptical about certain information being shared in OHCs: *I don't know if I can trust these online forums. It's people who just say, "Oh, I've tried this diet, it's not working for me." But how do you know they did it 100%, or if they are doing it wrong?* (P9)

Accordingly, Scientists need to see strong evidence to trust what other members are sharing. P4 noted the importance of "science-based" information that relates to generalizability over anecdotal evidence: *[I need] something that's more science based and relatable to the general public than just one person. [...] you often get a lot of things that aren't very reputable and are not really backed up by any science* (P4).

As cues for credibility, Scientists look for references, numbers, and recent postings (P8). They will also cross-validate multiple forums to validate information (P9). P9 did not trust posts with URLs in the end, because it might be links to promoting a product. Scientists will also look at whether a post is paragraphed or just a "giant wall of text" to assess the quality of the post. P9 saw that *"not necessarily people who wrote long text replies are good."*

Although these simple cues can get Scientists to detect the "scent" for quality, often participants had to read further to get a good sense of how valid the information being shared is. Scientists need to closely check who the posters are so that they can judge if the person is "flagged somewhere else working for whatever company" (P5). Also, presenting information "moderately, not extreme" (P8) would help gain readers' trust.

Unlike Coddlers, who read all the posts whenever possible, Scientists would stop reading the thread when members start to veer away from the main topic and become irrelevant (P6, P8).

Values sought: Finding others' experiences within spheres. As they make sure information is accurate, Scientists want to retrieve relevant information for their needs. They value others' experiences shared in OHCs. However, for these experiences to work for Scientists, they need to verify that these experiences are applicable to them. They have specific "spheres" they consider as important in finding the information they need. Spheres refer to areas of interest or constraints, such as money or primary health interests (e.g., losing weight or being heart-healthy) (P8). For instance, patients with type I diabetes patients will not want to read experiences of type II diabetes patients (P7). Similarly, for Scientists, it is important to know the posters and their profile in depth so that they find relevant others: *"This is what you need. You just need to eat that vs. this and this." But who did that? And what is that person's condition? And why didn't he do that? I need to learn more about why they did that and how did they do that (P9).* P5 also looked for a similar demographical profile when trying to learn about post-surgery management.

### *Adventurers: An exploring patient—OHC as a place to explore new and alternative ideas*

Unlike Scientists, Adventurers do not care about scientific evidence. They want to be challenged and are curious about how others talk about managing illness regardless of whether evidence exists. While Scientists frown upon information not validated by the scientific community, Adventurers value cutting-edge information on radical approaches with potentially better results that members discuss, which might not be embraced by the mainstream, conservative medical community. To be an adventurer, one needs to have substantial knowledge behind as a diabetes patient. Accordingly, adventurers tend to be experienced patients with having diabetes relatively under control, always open to exploring new and novel ways of managing the illness in better ways.

Participation: Messengers or steady observers. Similar to Scientists, Adventurers do not have a common participation pattern. Adventurers either become the messengers of new information, posting useful articles, or steadily come to OHCs to observe people's experiences. When browsing through OHCs, they do not necessarily know what they are looking for.

Because Adventurers are open to new ways of managing illness, sharing what worked for them, especially new regimen, is a good way to mutually benefit from participating in OHCs: *I try to share my results. This is what works for me, this is what I find that's beneficial to me. [...] I like being helped by other people that may have had similar instances. How to use supplements? What supplements they've used that have worked, that don't work* (P2). P2 also shares articles with the community if he runs into the information that might enlighten the community members.

Consumption: Curious and open exploration. Adventurers are curious and open to exploration. Most of the time they do not know what they are looking for. They do not care about finding relevant information for them. They want to be surprised and let unusual information catch their interests. For instance, P2 called himself as a person who says a lot of "unpopular things." In a way, Adventurers consider themselves as outsiders who are into extreme things: *I would be curious as to what somebody was saying that was so unpopular, because odds are pretty good they may be saying what I'm saying, because I say a lot of unpopular things* (P2).

Adventurers get excited when seeing individual differences: *Wow. Wow. That would be, that's an interesting post. It's really definitely one of those, "What works for me doesn't work for you."* (P2) This view of an Adventurer contrasts with Coddlers who feel uncomfortable when they see disagreeing facts (P10). P2 actually found disagreeing posts to be useful compared to "cheerleading" posts: I'm looking for someone who goes, "This is a bunch of baloney, and I tried this and it didn't work." Sometimes you can find more information from the negative reviews or negative comments than you can from the cheerleaders. (P2)

Adventurers' practice of consuming information in OHCs is exploratory. Adventurers rarely have a focus while browsing through OHCs. P11 found suggested keywords during search to be helpful when he does not know what he wants. Adventurers want to encounter unusual information that is out of the ordinary: *Things that are interesting are generally things that either pertinent to me or are sort of out of the ordinary* (P4).

Adventurers find it helpful to read those that challenge what they already know: *Because it doesn't fit in my mind of how it should work. I mean that doesn't mean that my mind is right, but it creates a conflict in my mind that I wanna try to figure out* (P3). Similarly, P1 often wants to "mix it up for a change" by randomly choosing different pages to read: *Or sometimes, I would do page, maybe one, two and three, and then go to page 10, 11, 12, something like that just to kind of mix it up and get further away from something just for a change* (P1).

Values sought: Avoiding the mainstream. Adventurers value OHCs for its potential to expose resources that doctors might have missed or refused to give. For Adventurers, OHCs are a place that provides diverse and cutting-edge information. Adventurers seek information that the medical community does not embrace due to its extremity. P2 believed that "most of the information that the established medical and diabetes community puts out there is very middle of the road." Thus, P2 valued OHCs for their exposures to "things on conversation groups that are outside the norm that are being provided by the established medical and diabetes community." Adventurers feel doctors do not share enough information with them or provide the information they want: *I am so frustrated that my doctor is no help, don't think seeing a diabetes educator is helpful* (P14).

Patients' experiences are another example that doctors cannot give to patients: *[The community has] mostly people who [has experiences with] either pre-surgery or post-surgery talking about their different experiences. You have a lot of stories on that forum* (P5). Some Adventurers do not necessarily choose to be an Adventurer but their personal context makes them become one: *online is the only place I can get any info. I also live in a very rural area, so going to the library is out of the question* (P14).

Adventurers, contrary to Scientists, appreciate contradictory, conflicting information. They care less about credibility as long as the information is novel and perceived as useful.

***Opportunists: Newly diagnosed--OHC is just one place out of many to get information***
Opportunists are not regulars or frequent visitors of an OHC like those with other personas. Opportunists land on one thread from an OHC by searching the Web. They get what they need from the thread and leave. They rarely stay to browse the community further; but if they do, they lurk. They consider OHCs as a place to get people's experiences for triangulating with other Web search results.

Participation: Landing user, lurkers. Newly diagnosed patients have many questions that they do not know where to find answers. They can be opportunists, who land on a thread of OHCs from a search portal. Opportunists are not registered in any forum (P8, P9). They lurk and leave (P8), or post one question at best. Opportunists do not go to other threads within the community after reading the thread—they come back to the list of search results (P6, P9, P12). Thus, Opportunists feel no ownership or social linkage with other community members. Opportunists browse through many search or curated search results to get to an OHC. When asked how the participant found the OHC thread she mentioned, P4 could not remember: *Oh, Jeez. How do you ever get to anything on reddit? Seeing it cross-posted or hearing people talk about it in a different sub reddit... Just browsing through, I guess* (P4).

Since Opportunists are not bound to one community, unlike Coddlers who find "everything under the sun" (P9) from one OHC, Opportunists like to triangulate the information found from an OHC with other information sources: *Once I make sure the information's correct, then I go back to the search results and then I just look for some blog or some online community that talks about the diet [from the] people who tried to do that diet* (P9).

Consumption: Skim and move on. Opportunists do not necessarily read all posts in depth. They get what they need and leave. They skim not only OHCs but also other websites as part of their information gathering process. P9 described a representative way of Opportunists' information search in general: *I go for a news article and see how many weights have he (a celebrity) lost, for how long, how many days or months. [...] I will jump to the next [search] results and see if it's more related to me (P9)*.

During P9's information search process, OHCs might or might not be included. For instance, P8 described how she travels in and out of OHCs through unguided browsing: *a friend will put something on Facebook "I just ran a marathon, read about it on my blog," And then I'll go to her blog, and then someone on the blog will comment about, "Oh, I had a similar experience. Come read about this at my Tumblr page or at this group on this website". And so, I'll just follow the electronic breadcrumbs and end up reading about other people's experiences* (P8).

In favor of fast browsing, they also care about the cost of reading, such as time and effort, when opening up a thread. They look for specific things in the posts: *I don't want to read all the texts and stuff. I'm looking for intersections of specific things. I don't want to spend a lot of times reading posts and see which posts will fit for me* (P9).

Opportunists look for numbers, signaling words, or pictures (P8) to skim and read only those posts that catch their interests or relevant to them. P4 described her recommendations on how to make skimming efficient: *I think that maybe two lines of information in the search result to further read more would be more useful, so that you could have a better idea of whether the rest of the post is going to be pertinent* (P4).

Values sought: Product reviews and experiences. Although Opportunists find it difficult to efficiently find factual, practical information, they see OHCs as a place to get people's experiences as additional information. For instance, P12 saw OHCs as not offering practical advice: *[One site I found] was a bunch of people talking back and forth and not really offering any real advice, I guess* (P12).

They find OHCs as a place to get "reviews," similar to Amazon product reviews (P6). For instance, P6 and P9 describe what he would use OHC for—as a place to get individuals' experiences on a regimen: *The only time I would consider reading [OHCs] would be after I feel I have a background [from] diabetics associations or whatever, and then consider reading this for individual people's experiences* (P6); *I Google-ed their diet and the forum to see if normal people like me did this and was working for them* (P9). Similarly, P7 described how she would triangulate resources from Google search with what she finds in OHCs, sometimes by asking questions: *if I can't find anybody on the forum that tried it, I can ask, "Has anybody tried this? What were your experiences with it? How did you like it?"* (P7)

***Multiple and transitioning nature of personas, intertwined with illness***
The four personas described so far might be somewhat artificial, stereotyping, and simplifying the behaviors and the views of our participants regarding OHC use. Rather than mutually exclusive personas, these personas should be viewed as types of participation that OHC users could engage in depending on their illness stage, personality, or support needs at any given time. Accordingly, at any point, an OHC user could possess multiple personas and move from one persona to another over time. Persona development generally can follow their illness trajectory, starting with newly diagnosed, to exploring patient, and to experienced patient. However, based on the needs at the time, the stage of illness the user is on will not immediately categorize an OHC user to one persona, due to the complicating situational factors that all play in their support needs at the moment.

Multiple nature of personas. Most of our participants showed multiple personas with one or two main personas. For instance, P2 was an experienced diabetes patient who possessed the characteristic of a Scientist needing scientific evidence for information he gathers. At the same time, he was also an Adventurer in that he wanted to explore new regimen whenever available. He showed a partial characteristic of a Coddler, in that he was altruistic to other community members, but being against coddling in OHCs. He believes people need to face the hard facts: *This is my complaint on discussion groups. People tend to coddle diabetics a lot because we have emotional issues, and we can be somewhat fragile* (P2).

At the same time, he liked to share his experiences with other people who are struggling, which is a characteristic of a Coddler. Another example is P3, who was a Coddler, a Scientist, and a little bit of an Adventurer. She regularly visited an OHC to interact with other people. She also liked to validate what other members had posted. She was adventurous in that she was interested in seeing posts in conflict with her knowledge.

Transitional nature of personas. The participants who were old-timers in OHCs and those who have had diabetes for a long time shared how their perspectives towards OHCs changed over time. These changing perspectives meant that their personas also moved from one to another. For instance, the participants we described as Coddlers all started out as landing users, meaning that they were opportunistically using OHCs at first: *I didn't even start talking to anybody until about two years ago, or responding to any discussions* (P10).

Over time, P10 started talking with other members and began socializing in the community as she began to regularly visit the community, similar to P3. P1 also confessed she never considered herself as being able to mutually exchange social support with others online and have healthier lifestyle as a consequence: *I didn't think I ever, ever could. I never thought I would be one of those people* (P1).

Personas intertwined with illness. Most of our participants did not consider the diagnosis as serious, thus waited for many years until they finally encountered life-threatening events (P1), such as a stroke (P2, P10) or frequent hospital visits leading to expensive medical bills (P3, P7, P14). They then began visiting and participating in OHCs to be surrounded by similar others (P9), gain education on their own (P3) and moved forward. An Opportunist can begin to stay in the community, participate as a moderator, and continue to do further research on what illness management regimen can be improved, evolving into a Scientist (P2).

Depending on their illness stages, user needs will be different. At the time of the diagnosis, patients have big questions (P9), exploring options as they attempt to find their own strategies that work best for them. P11 showed frustrations for not being able to get "straight up" answers. Similarly, P6 did not feel OHCs address the "facts well." However, P9 also initially had such informational needs only, but over time started to socialize with others and found value in mutual support in the communities. As patients get settled onto their routines, they might become more skeptical or more open to new kinds of information. They could develop frustration over an existing regimen as it no longer works. Thus, they might turn to alternative ways of managing illness like Adventurers.

### *The perspectives and needs of the OHC administrator*

According to the OHC administrators we talked to, the goal in managing online health communities is to make sure: (1) those who need help gets help, (2) monitor conversations that can be destructive to the community, (3) understand patients' illness stage and provide appropriate help accordingly, and (4) profile members to provide personalized help. These four tasks are done manually by reading the community threads everyday and monitoring them closely. Accordingly, some automated ways to detect these notable activities would be helpful for them. When presented with the personas and their illness contexts, the OHC administrators agreed that the personas described well the patient profiles they had in mind as they moderated their OHCs. Furthermore, the OHC administrator noted that the majority of the community members would fall into the category of Opportunists—the lurkers—many of whom might need help, due to their newly diagnosed state.

## Discussion

In this section, we discuss how our findings might apply to other illness contexts. We then discuss how our work contributes to existing online community and information behavior literature. We note limitations of our study, and end with requirements for delivering personalized, large-scale online conversations in OHCs.

The four personas in other illness contexts. We generated our findings in the context of diabetes. Whether our findings apply to other illness contexts is an open question. Depending on the illness, one persona might be more popular than others. Moreover, the behavioral characteristics grouped into the four personas might not be the same for other illness contexts. As Huh et al. [28] found, Youtube users with cancer exchanged more emotional support than those with diabetes, who exchanged mainly informational support. Thus, for those with cancer, Coddlers could make up the most user population in a given OHC. These Coddlers with cancer might have more characteristics of Adventurers in them than our participants—certain cancers do not have known effective treatments, and thus, people might turn to alternative ways of finding appropriate remedies.

Connecting to online communities and information behavior literature. Our work showed that OHC personas are fluid, complex, and boundless, challenging existing static and deterministic models towards social types in online communities and information seeking literature. We saw how Coddlers vs. Adventurers show differences in their attitudes in dealing with conflicting information. Coddlers tend to turn away or ignore extreme, contradictory information. On the other hand, Adventurers look for the information that could challenge them. Such attitude differences resonate with what Steptoe and O'Sullivan [29] described as blunters and monitors. Blunters turn away from, but monitors face any conflicting information. However, our personas describe how blunting and monitoring are fleeting activities, always ready to change depending on the context of the illness and how other community members interact with each other. After all, both blunting and monitoring are ways to ameliorate cognitive dissonance [30] and selective exposure [31] one way or the other.

Kim [32] described social types according to how one gains participation level with the community over time. We did observe some of our participants walking that path that Kim had described—and eventually transitioning their personas from Opportunists to Coddlers. However, such transitions are not simply based on a participation level with the community. The transition involves moving forward with their illness management, stages of behavioral change, and evolution of interpersonal relationships with other community members over time.

We do not see that our four personas are complete or sufficient of all OHC users' behaviors. We had participants who are weakly associated with multiple personas. How exclusive each persona is should be carefully assessed when applying these personas for other OHC contexts. Such multiplicity of our personas agree with Zhao et al. [33]'s work on maintaining multiple faces on Facebook; Also, Goffman's presentation of self [34] and Laurel's virtual self [35] all point to similar lines of argument that people exhibit varying self online. The unique contribution of our personas, however, is that we: (1) identified clusters of information seeking and online participation behaviors and (2) began to illustrate patterns for how illness contexts, personal beliefs, and social context shape the personas.

Design implications. We see three main areas in which our findings can inform future design of OHCs. **(1) Persona/participant type Detection:** Systems can profile each user's interests towards psychosocial and informational support over time. Hartzler et al.[36] used past posts of OHC users to profile their health interests. From these health-related profiles, we can further examine users' dominant persona at any given time and provide appropriate metrics to help them explore through OHC content. **(2) Quality metrics development:** Our participants used cues (e.g., references, numbers, other peoples' confirmations) to assess the qualities of posts. It would help if OHCs can index and present each post with enriched cues using metadata corresponding to these qualities. Systems can show the strengths of each poster's expertise by assessing their participation patterns, contents of posts, and reputation from other members. Users can freely create their own virtual "spheres" (e.g., money saving, vegan, pregnant, married, etc). Users can drag threads into these "spheres" that belong together. Recommendation systems can suggest posts to read and people to connect with, using the quality metrics for each individual depending on their predicted persona. **(3) Information scent and Recommendation.** For OHC search results, design features, such as glyph icons[37] can provide information scent around dominant persona of the poster, quality metrics, and trails of what kinds of personas have visited the thread. These features can be used to effectively summarize the post content. Small glyphs can be useful for Opportunists and Adventurers who often skim through search results to explore 'something interesting'.

**Conclusions**

In this paper, we identified four personas that help illustrate diverging user needs in OHCs. These personas helped us understand the fluid nature of an illness context, which drives online community participation behavior. Our work contributes to: (1) identifying the unique psychosocial and dynamic nature of personas in OHCs; (2) extending online community user and information behavior literature by adding how illness contexts dynamically shape information behavior; and (3) extracting requirements for helping users consume large-scale content in OHCs in a personalized manner. Our participants all had their own ways of coping with challenges—through science, psychosocial support, adventure, or being opportunistic. Our study is a stepping stone to supporting these diverging perspectives for strong, personalized health support.